\documentclass[pra,twocolumn,showpacs]{revtex4-1}

\usepackage[SquareTraceBrackets]{quantum}
\usepackage{graphicx,bm,natbib,upgreek,amsbsy,mathrsfs,amsfonts}

\renewcommand{\Re}{\mathrm{Re}}
\renewcommand{\Im}{\mathrm{Im}}

\DeclareMathOperator{\lcm}{LCM}

\begin{document}

\title{\large\textsf{\textbf{Unitary evolution and the distinguishability of quantum states}}}

\author{Sam Morley-Short}
\author{Lawrence Rosenfeld}
\author{Pieter Kok}\email{p.kok@sheffield.ac.uk}

\affiliation{Department of Physics \& Astronomy, University of Sheffield, Sheffield S3 7RH, United Kingdom}

\begin{abstract}
 \noindent The study of quantum systems evolving from initial states to distinguishable, orthogonal final states is important for information processing applications such as quantum computing and quantum metrology. However, for most unitary evolutions and initial states the system does not evolve to an orthogonal quantum state. Here, we ask what proportion of quantum states evolves to \emph{nearly} orthogonal systems as a function of the dimensionality of the Hilbert space of the system, and numerically study the evolution of quantum states in low-dimensional Hilbert spaces. We find that, as well as the speed of dynamical evolution, the level of maximum distinguishability depends critically on the Hamiltonian of the system. 
\end{abstract}

\date{\today}

\maketitle

\section{Introduction}\noindent
A question of both fundamental and practical interest in quantum mechanics is how fast systems evolve from an initial state to an orthogonal final state \cite{margolus98,soderholm99,giovannetti03,jones10,zwierz12,taddei13,delcampo13,margolus14}. The practical importance is due to the perfect distinguishability of orthogonal states in single-shot measurements, and these play a crucial role in metrological applications \cite{giovannetti06,zwierz10,giovannetti11}. From a fundamental perspective, the dynamical speed of evolution can be used to prove uncertainty relations \cite{mandelstam45,bhatta83,uffink93}. However, for non-interacting finite-dimensional systems, the set of states that ever reach an orthogonal state via free evolution has measure zero, and results that rely on strict orthogonality can only be approximately true. 

Here, we study how quantum systems of various dimensions evolve to their most distinguishable state. We consider two classes of Hamiltonians that are defined by their energy spectrum, and study how they lead to nearly distinguishable states. Lack of orthogonality has been studied before, including its effect on bounds \cite{soderholm99,giovannetti03}. However, the precise dynamics of quantum systems has not been studied in detail, and in particular it is not known how rapidly systems achieve near-orthogonality. Here, we provide a numerical answer to this question by uniformly sampling the state space in dimensions $N=2$ to $N=20$. While it is well-known that the speed of dynamical evolution depends on the dynamics of the system, we find that the average maximum attained distinguishability also depends on the details of the dynamics. 

In Sec.~\ref{sec:dist} we review the definition of distinguishability and set up the problem. In Sec.~\ref{sec:ham} we describe the two classes of dynamical systems under consideration, and in Sec.~\ref{sec:qsl} we discuss the implications of our results on quantum speed limits. Finally, in Sec.~\ref{sec:con} we present our conclusions.

\section{Distinguishability of quantum states}\label{sec:dist}\noindent
Two quantum states $\ket{\psi}$ and $\ket{\phi}$ are perfectly distinguishable in a single measurement when their inner product $\braket{\phi|\psi}$ is zero. The measured observable can then be chosen such that the states $\ket{\psi}$ and $\ket{\phi}$ are eigenstates with different measurement outcomes (i.e., the physical eigenvalues). It is well known that the absolute square of the inner product is the \emph{fidelity} for pure quantum states, which can be interpreted as the probability of mistaking one state for the other in a single-shot measurement:
\begin{align}
 F = \abs{\braket{\phi|\psi}}^2\, .
\end{align}
This immediately suggests a continuous scale for the \emph{distinguishability} of the two states as $1-F$ \cite{giovannetti03}.

Next, we consider an $N$-dimensional isolated quantum system $S$ described by a Hamiltonian $H$. The system is in a quantum state $\ket{\psi}$. After a time $t$, the state will have evolved to 
\begin{align}
 \ket{\psi(t)} = U(t) \ket{\psi} = \exp\left(-\frac{i}{\hbar} H t \right) \ket{\psi}\, ,
\end{align}
where $U(t)$ is the free unitary evolution for a duration $t$. The distinguishability is then easily calculated as $1-\abs{\braket{\psi|\psi(t)}}^2$. However, we are really interested in the maximum distinguishability between $\ket{\psi}$ and $\ket{\psi(t)}$. The states $\ket{\psi(t)}$ form closed orbits in the state space of $S$ with period $T$ (due to the quantum mechanical version of Poincar\'e's recurrence theorem), which means that there is a time $\tau < T$ that minimises $\abs{\braket{\psi|\psi(t)}}^2$. This leads to the concept of the \emph{maximum distinguishability} $D$:
\begin{align}
 D \equiv 1-\abs{\braket{\psi|\psi(\tau)}}^2\, .
\end{align}
This definition is readily extended to the evolution of mixed states by using the Uhlmann fidelity \cite{uhlmann76,jozsa94}. Here we will restrict ourselves to pure states and unitary evolutions, since it is the most fundamental quantum mechanical situation.

To further argue that $D$ is a natural measure of distinguishability, we show that this quantity behaves correctly for composite systems. Each composite system of two systems with dimensions $N_1$ and $N_2$ in states $\ket{\phi}$ and $\ket{\psi}$, respectively, can be written as a single system in the state $\ket{\Psi} = \ket{\phi}\otimes \ket{\psi}$ with dimension $N_1 \times N_2$. Suppose that the state $\ket{\phi}$ does not evolve in time. Then the maximum distinguishability is entirely determined by the state $\ket{\psi(t)}$ and we find 
\begin{align}
 D_2 = 1 - \abs{\braket{\psi|\psi(t)}}^2\, .
\end{align}
Alternatively, we can calculate the $D$ for the composite system, which yields 
\begin{align}
 D_{12} = 1 - \abs{\braket{\Psi|\Psi(t)}}^2 =  1 - \abs{\braket{\phi|\phi}}^2 \abs{\braket{\psi|\psi(t)}}^2\, .
\end{align}
Since $\abs{\braket{\phi|\phi}}^2=1$ we find that $D_{12} = D_2$, and the distinguishability behaves as one would expect.

To calculate the maximum distinguishability of arbitrary quantum states we must choose a representation that is easily implemented numerically. 
A general quantum state in $N$ dimensions can be written as 
\begin{align}\label{eq:inputstate}
 \ket{\psi} = \frac{1}{r}
 \begin{pmatrix}
  e^{i\varphi_1} x_1 \\ e^{i\varphi_2} x_2 \\ \vdots \\e^{i\varphi_{N-1}} x_{N-1} \\ x_N 
 \end{pmatrix}\, ,
\end{align}
where the $x_i$ are cartesian coordinates and $e^{i\varphi_j}$ are complex phases. The factor $1/r$ ensures that $\ket{\psi}$ is normalised, with $r^2 = \sum_{i=1}^N x_i^2$. We can assume that the Hamiltonian of the system is diagonal in the basis implied by Eq.~(\ref{eq:inputstate}) without loss of generality, since $\ket{\psi}$ itself is completely arbitrary. After free evolution for a time $t$, the state then evolves into
\begin{align}
 \ket{\psi(t)} = \frac{1}{r}
 \begin{pmatrix}
  e^{i\varphi_1-i\omega_1 t} x_1 \\ e^{i\varphi_2-i\omega_2 t} x_2 \\ \vdots \\e^{i\varphi_{N-1}-i\omega_{N-1} t} x_{N-1} \\ e^{-i\omega_N t}x_N 
 \end{pmatrix}\, ,
\end{align}
where $\omega_j = E_j/\hbar$, and $E_j$ are the eigenvalues of $H$. The maximum distinguishability for the input state $\ket{\psi}$ can then be written as 
\begin{align}\label{eq:dist}
 D = 1 - \frac{1}{r^4}\Abs{\sum_{n=1}^N e^{-i\omega_n \tau} x_n^2}^2\, .
\end{align}
Note that $D$ does not depend on the initial phases $\varphi_j$ of the quantum state at all.

Note that it is extremely unlikely that any given $\ket{\psi}$ evolves to an orthogonal state. For $N=2$ this is obvious: choose $\ket{\psi}$ somewhere in the $xz$-plane of the Bloch sphere, and assume without loss of generality that the Hamiltonian is proportional to the Pauli matrix $\sigma_z$. The state will evolve to an orthogonal state only if it lies in the equatorial $xy$-plane, perpendicular to the $z$-axis. The set of states that evolve to an orthogonal state lie on a one-dimensional line (the equator), while the totality of states is described by a two-dimensional surface. The set of states that evolve to orthogonal states therefore has measure zero with respect to the entire state space. This behaviour persists in higher dimensions. To speak meaningfully about distinguishability, we therefore introduce a parameter $\epsilon$, the value of which must be determined by external factors (such as precision requirements, fault tolerance thresholds, etc.), and that indicates a minimum distinguishability. In other words, we consider the probability that an input state $\ket{\psi}$ achieves a distinguishability of $D \geq 1-\epsilon$. This allows us to study the evolution of quantum states as a function of the dimension $N$ of the system, and speak of \emph{near-orthogonality} in a meaningful way. 

\begin{figure}[t!]
\includegraphics[width=7cm]{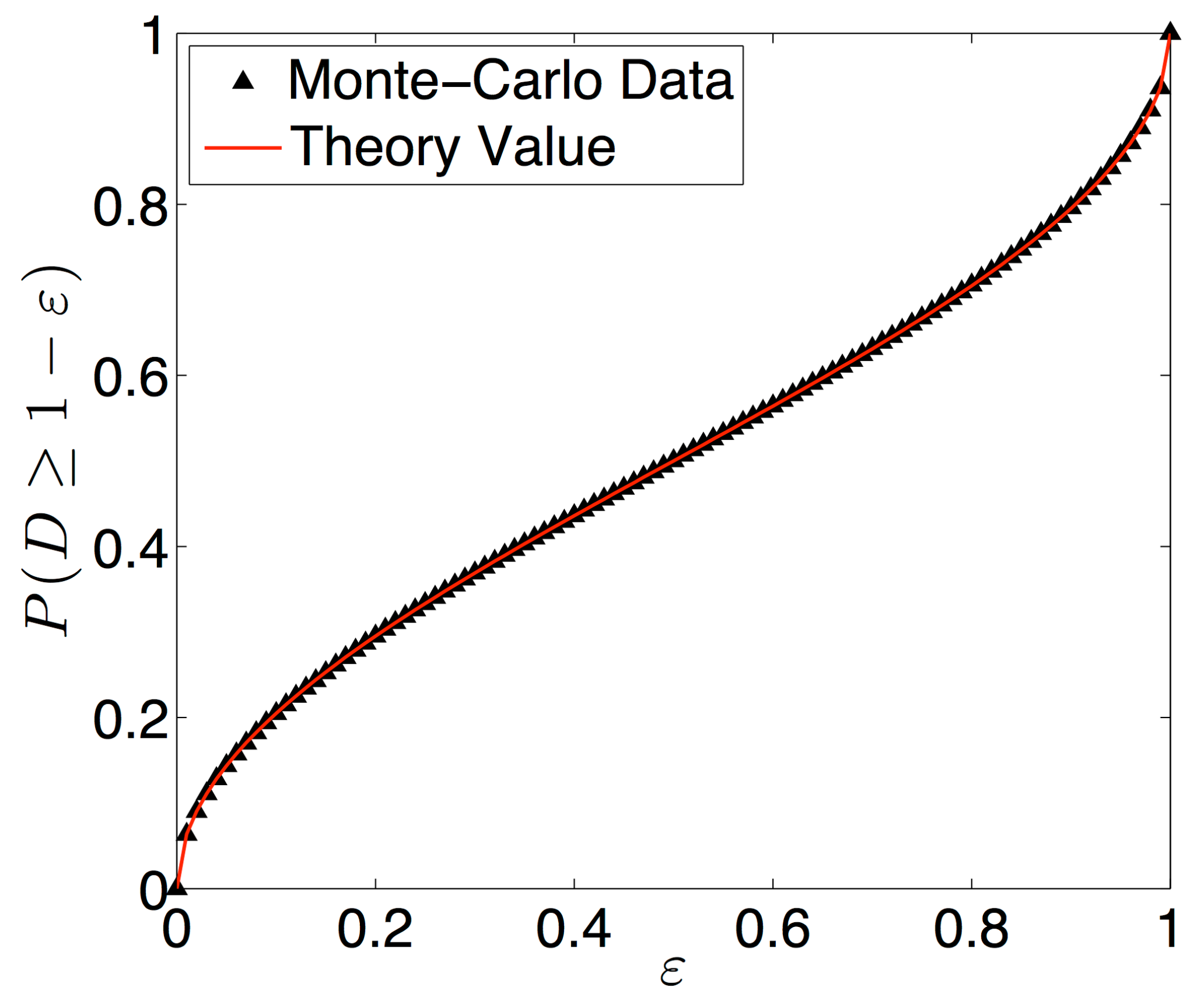}
\caption{(color online) The probability that a randomly chosen state $\ket{\psi}$ evolves to a state $\ket{\psi(t)}$ with maximum distinguishability $D\geq 1-\epsilon$. The solid curve is the theoretical value of Eq.~(\ref{eq:n2eps}), while the triangles are obtained by a Monte Carlo approach.}
\label{fig:n2eps}
\end{figure}

Again for the case of $N=2$, we can calculate the probability that a randomly chosen state $\ket{\psi}$ evolves to a state $\ket{\psi(t)}$ with maximum distinguishability $D\geq 1-\epsilon$. In appendix \ref{app:pop} we prove that 
\begin{align}\label{eq:n2eps}
 \pr{D\geq 1-\epsilon} = \frac{2}{\pi} \left( \arctan \alpha_+ - \arctan \alpha_- \right)\, ,
\end{align}
where 
\begin{align}
 \alpha_\pm = \sqrt{ \frac{2}{1\pm\sqrt{\epsilon}}-1 }\, .
\end{align}
See also Ref.~\cite{giovannetti03}.
This probability distribution as a function of $\epsilon$ is shown in Fig.~\ref{fig:n2eps}, along with numerical values that we obtained using a Monte Carlo approach. In higher dimensions the analytic solutions become intractable, and we must rely on numerical simulations alone.

Next, we will explore the maximum distinguishability attained by quantum states in dimensions up to $N=20$ via Monte Carlo simulations. We uniformly sample the quantum state space---i.e., the values of $x_j$ in Eq.~(\ref{eq:inputstate})---and calculate the time $\tau$ that maximises $D$. The populations of different $D$ are plotted in histograms, which allows us to see straight away how the maximum distinguishability changes with $N$. However, before we can present these results, we first have to consider the Hamiltonians of the systems.

\section{Hamiltonians}\label{sec:ham}\noindent
The maximum distinguishability in Eq.~(\ref{eq:dist}) depends on the energy eigenvalues $\omega_n$ of $H$ (up to a factor $\hbar$). This means that different Hamiltonians will generally lead to different behaviour in attaining certain levels of distinguishability. To study these differences, we consider two classes of Hamiltonians, which we term ``harmonic'' and ``atomic''. These Hamiltonian classes are motivated by their physical relevance: harmonic Hamiltonians have equally spaced energy values:
\begin{align}\label{eq:harmonic}
 \omega_n = n \omega \qquad\text{with}\qquad n\in\{1,\ldots,N\}\, .
\end{align}
Here, $\omega$ can be any angular frequency.  

By contrast, atomic Hamiltonians have large variety in the spacing between the energy levels. We choose a truncated version of the Bohr model for our atomic Hamiltonians:
\begin{align}\label{eq:atomic}
 \omega_n = - \frac{\omega}{n^2} \qquad\text{with}\qquad n\in\{1,\ldots,N\}\, .
\end{align}
The latter class of Hamiltonians has decreasing energy gaps as $n$ increases, and may seem restrictive. However, due to the invariance under relabelling of the basis vectors in the Monte Carlo procedure, this class includes all Hamiltonians that have a $n^{-2}$ absorption spectrum. We could have chosen higher powers of $n$ to make the distinction more extreme, but we already find significant differences from the harmonic Hamiltonians using this physically motivated atomic Hamiltonian.
 
\begin{figure}[t!]
\includegraphics[width=8.5cm]{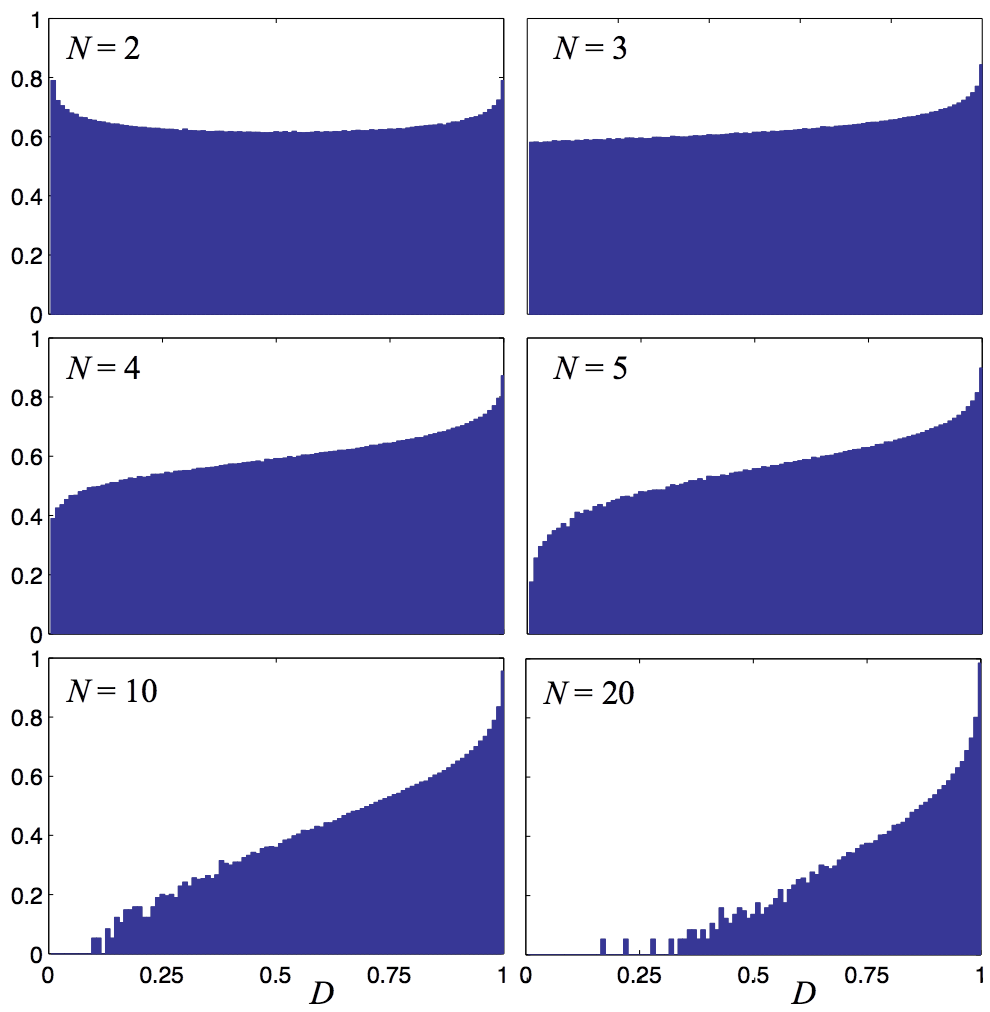}
\caption{Populations of different maximum distinguishability $D$ for representative dimensions (logarithmic scale) of systems with a harmonic Hamiltonian. The plots are normalised such that the total population is 1. The bin size $\Delta D$ is $0.01$. Note the inflection point occurring at $N=4$.}
\label{fig:ho}
\end{figure}

\begin{figure}[t!]
\includegraphics[width=8.5cm]{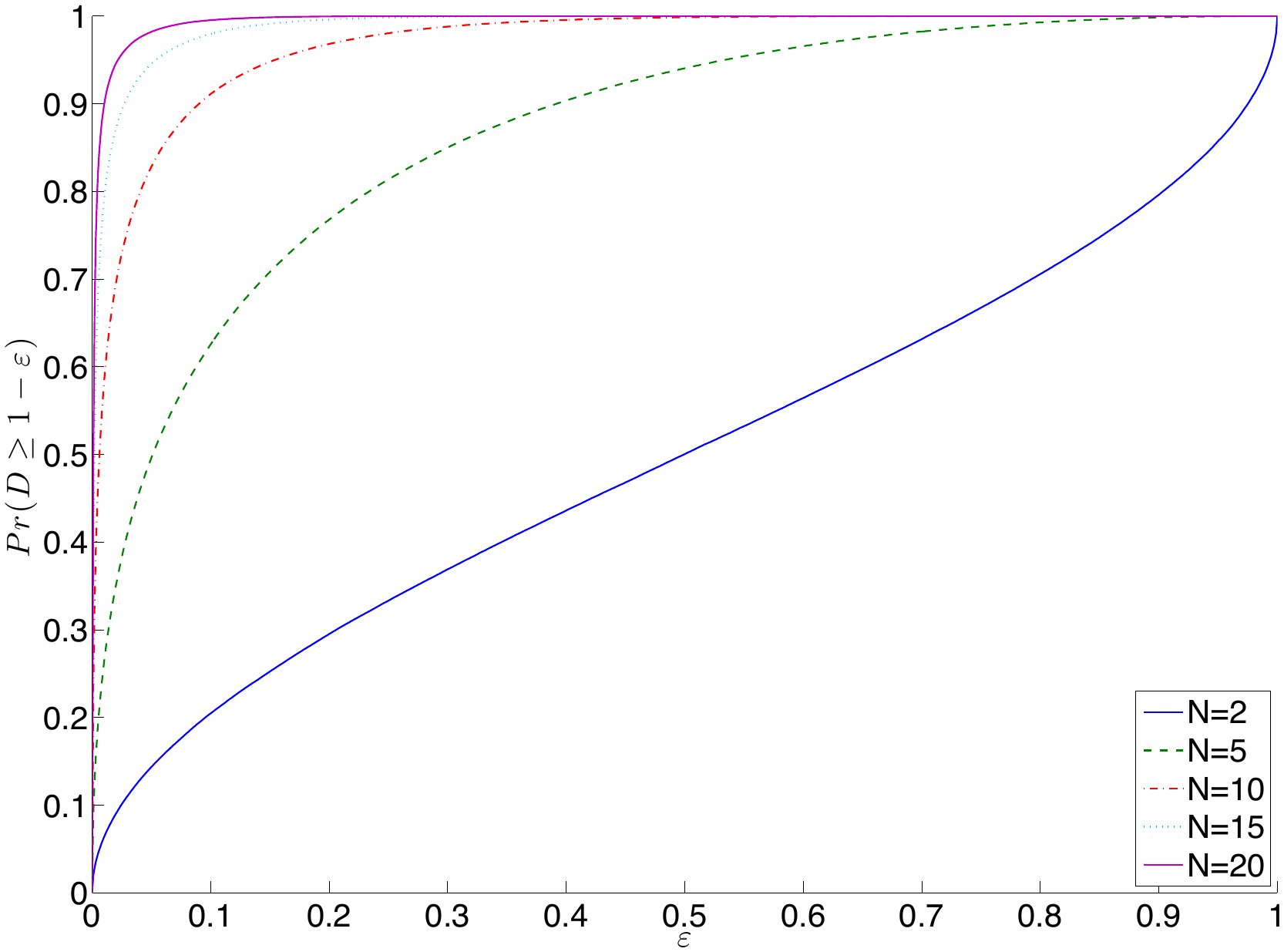}
\caption{(color online) The probability of picking a state with maximum distinguishability $D \geq 1-\epsilon$ as a function of $\epsilon$ for different dimensions $N$ with a harmonic Hamiltonian.}
\label{fig:pdho}
\end{figure}

\subsection{Harmonic Hamiltonians}\noindent
Our first task when evaluating $D$ is to find the minimum time $\tau$. To this end, we substitute Eq.~(\ref{eq:harmonic}) into Eq.~(\ref{eq:dist}) and find
\begin{align}
 D & = 1- \frac{1}{r^4} \Abs{\sum_{n=1}^N x_n^2 e^{-in\omega t}}^2 \cr
 & = 1- \frac{1}{r^4} \sum_{n,m=1}^N x_n^2 x_m^2 e^{-i(n-m)\omega t}
\end{align}
For simplicity we substitute $p_n \equiv x_n^2/r^2$, which are real numbers between 0 and 1, and $\sum_n p_n = 1$ (i.e., they are probabilities). This leads to the expression
\begin{align}
 D = 1- \sum_{n=1}^N p_n^2 - 2\sum_{n>m} p_n p_m \cos[(n-m)\omega t]\, ,
\end{align}
which is a periodic function with period $T=2\pi/\omega$. To find the extrema of $D$ we evaluate the derivative of $D$ with respect to time and find
 \begin{align}
 \frac{dD}{dt} = 2\omega \sum_{n>m} p_n p_m (n-m) \sin[(n-m)\omega t] = 0\, . 
\end{align}
Since the $p_n$ are non-negative and $(n-m)$ is positive, an extremum in $D$ will occur when for non-zero $p_n$ and $p_m$ the factor $\sin[(n-m)\omega t]$ is zero, or
\begin{align}
 \omega t = \frac{\pi}{n-m} + k_{nm}\pi \qquad\text{for all}~ n,m\, ,
\end{align}
where $k_{nm}$ is an integer that must be chosen such that all $\sin[(n-m)\omega t]=0$.
The shortest time to the maximum distinguishability is therefore given by states that maximise $(n-m)$---in other words, superpositions of states with the lowest and the highest energy eigenvalues. Moreover, when these states have equal amplitude, the system evolves to an orthogonal state. This is consistent with previous findings \cite{margolus98,levitin09}. For arbitrary states the time $\tau$ that maximises the distinguishability lies in the interval
\begin{align}
 \frac{\pi}{\omega} \frac{1}{N-1} \leq \tau \leq \frac{\pi}{\omega}\, .
\end{align}

In general, the sinusoidal modulations that need to be chosen zero depend on the values of $p_n$, and are therefore determined by the starting state $\ket{\psi}$. This is implemented as part of the Monte Carlo simulation, and in Fig.~\ref{fig:ho} we present histograms for the maximum distinguishability of $10^6$ uniformly sampled states for a system with a harmonic Hamiltonian. The histograms are plotted on a logarithmic scale and are normalised such that the population in the bin $D=1$ is 1. For higher dimensions, the populations become highly skewed towards higher maximum distinguishability, as expected.

The Monte Carlo data also allows us to chart the probability of picking a state with maximum distinguishability greater than $1-\epsilon$, which is shown in Fig.~\ref{fig:pdho}. In low dimensions, the evolution to a near-orthogonal state is indeed very unlikely. However, depending on the requirements on $\epsilon$, modestly sized systems (e.g., $N=20$) do have a very good chance of evolving to near-orthogonal states.

\begin{figure}[t!]
\includegraphics[width=8.5cm]{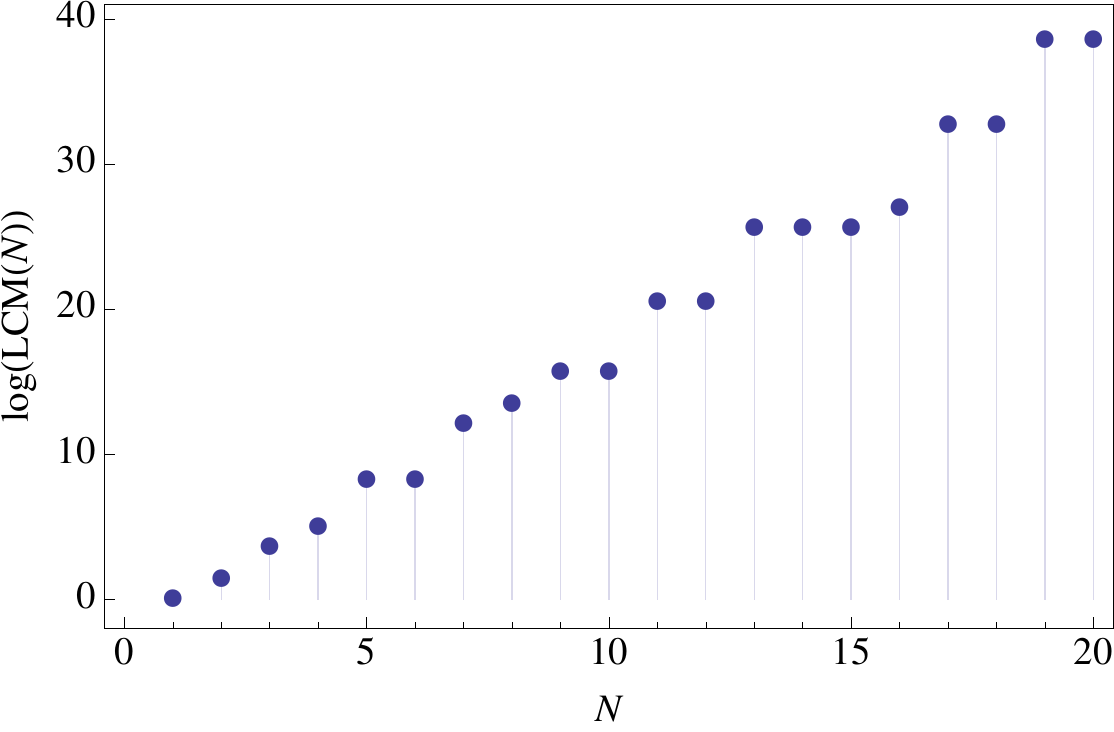}
\caption{Lowest Common Multiple of the set $\{ 1^2, 2^2,\ldots,N^2 \}$ (logarithmic scale).}
\label{fig:lcm}
\end{figure}

\begin{figure}[t!]
\includegraphics[width=8.5cm]{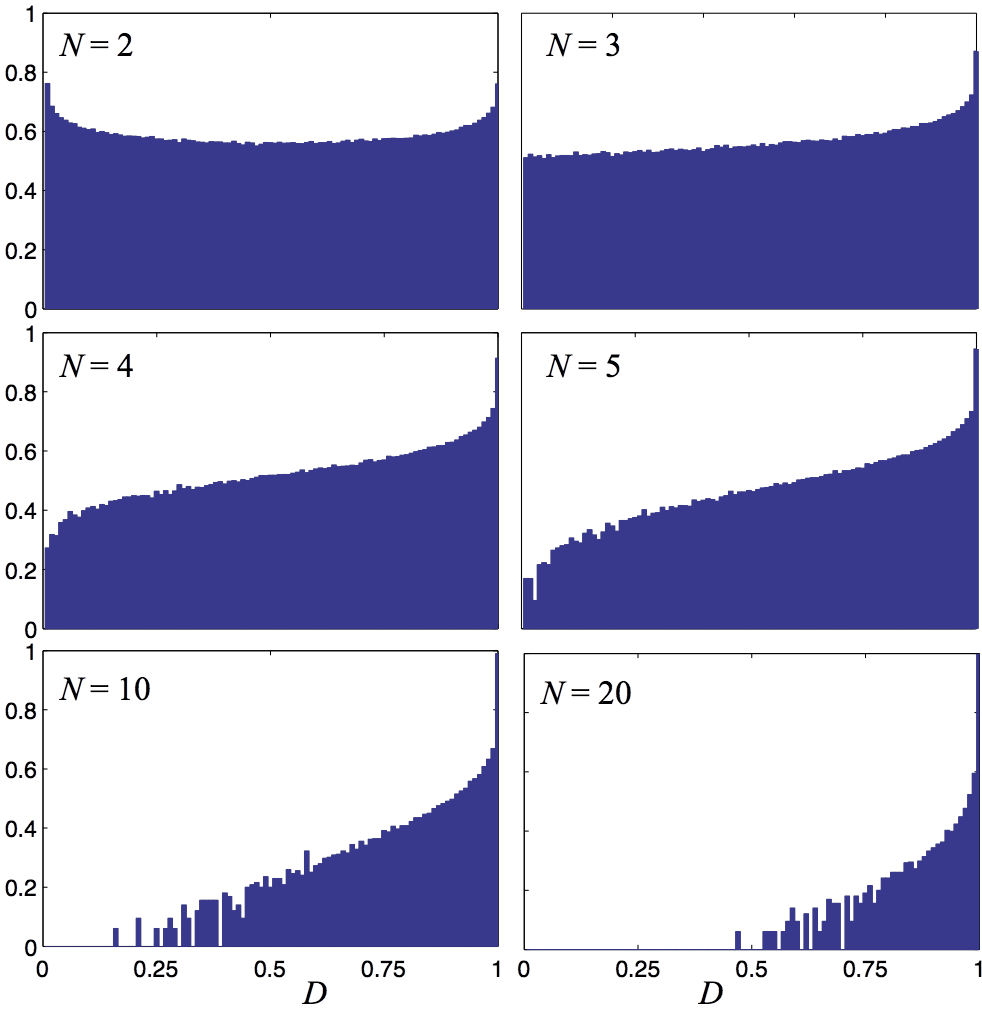}
\caption{Populations of different maximum distinguishability $D$ for representative dimensions (logarithmic scale) of systems with an atomic Hamiltonian. The plots are normalised such that the population in the bin $D=1$ is 1. The bin size $\Delta D$ is $0.01$. The inflection point occurs again at $N=4$, but states achieve near-orthogonality significantly faster compared to harmonic systems.}
\label{fig:atom}
\end{figure}

\subsection{Atomic Hamiltonians}\noindent
We repeat the procedure of the previous section for the class of atomic Hamiltonians of the form 
\begin{align}
 \omega_n = - \frac{\omega}{n^2} \qquad\text{with}\qquad n\in\{1,\ldots,N\}\, .
\end{align}
The extrema of $D$ occur when 
 \begin{align}\label{eq:atomicderivat}
 \frac{dD}{dt} = 2\omega \sum_{n>m} p_n p_m \frac{n^2 - m^2}{n^2 m^2} \sin\left[\left(\frac{n^2 - m^2}{n^2 m^2}\right)\omega t\right] = 0\, . 
\end{align}
The solutions with the shortest periods are again those with contributions from only the lowest and highest energy eigenvalues, leading to a minimum time 
 \begin{align}
 \tau = \frac{\pi}{\omega} \frac{N^2}{N^2-1}\, . 
\end{align}
However, for superpositions with nearly all $p_n$ non-zero we require that 
\begin{align}
 \omega t = \frac{ n^2 m^2}{n^2-m^2}\, \pi+ l_{nm}\pi \qquad \text{for all}~n,m\, ,
\end{align}
where $l_{mn}$ is an integer that must be chosen such that $\sin[(n^2-m^2)\omega t /(n^2 m^2)]=0$. Since the ratio $(n^2-m^2)/(n^2 m^2)$ is in general not an integer, different terms in Eq.~(\ref{eq:atomicderivat}) can have periods that are very close together. Consequently, the overall period of Eq.~(\ref{eq:atomicderivat}) grows rapidly with the dimension of the system $N$, and is given by the Least Common Multiple (LCM) over the set $\{ 1^2, 2^2,\ldots,N^2 \}$. Our numerical search for $\tau$ is then restricted to the interval
 \begin{align}
 \frac{\pi}{\omega} \frac{N^2}{N^2-1} \leq \tau \leq \frac{\pi}{\omega} \lcm(1^2,2^2,\ldots,N^2)\, .
\end{align}
The logarithm of the LCM for the set $\{ 1^2, 2^2,\ldots,N^2 \}$ is shown in Fig.~\ref{fig:lcm}. Clearly, the LCM increases exponentially, and the determination of the maximum distinguishability for atomic systems is computationally harder than for harmonic systems.

We sampled the quantum state space $10^6$ times and calculated $\tau$ and $D$. The results are again histograms of populations for all dimensions from $N=2$ to $N=20$. Representative dimensions are shown in Fig.~\ref{fig:atom}. It is clear that for moderate dimensionality ($N= 10$ to $N=20$) the atomic systems are much more likely to achieve near-orthogonality than the harmonic systems. This is somewhat surprising, since one could have expected that the difference between the two types of systems would manifest itself mainly in the speed at which it achieves near-orthogonality, not the \emph{level} of orthogonality.

\begin{figure}[t!]
\includegraphics[width=8.5cm]{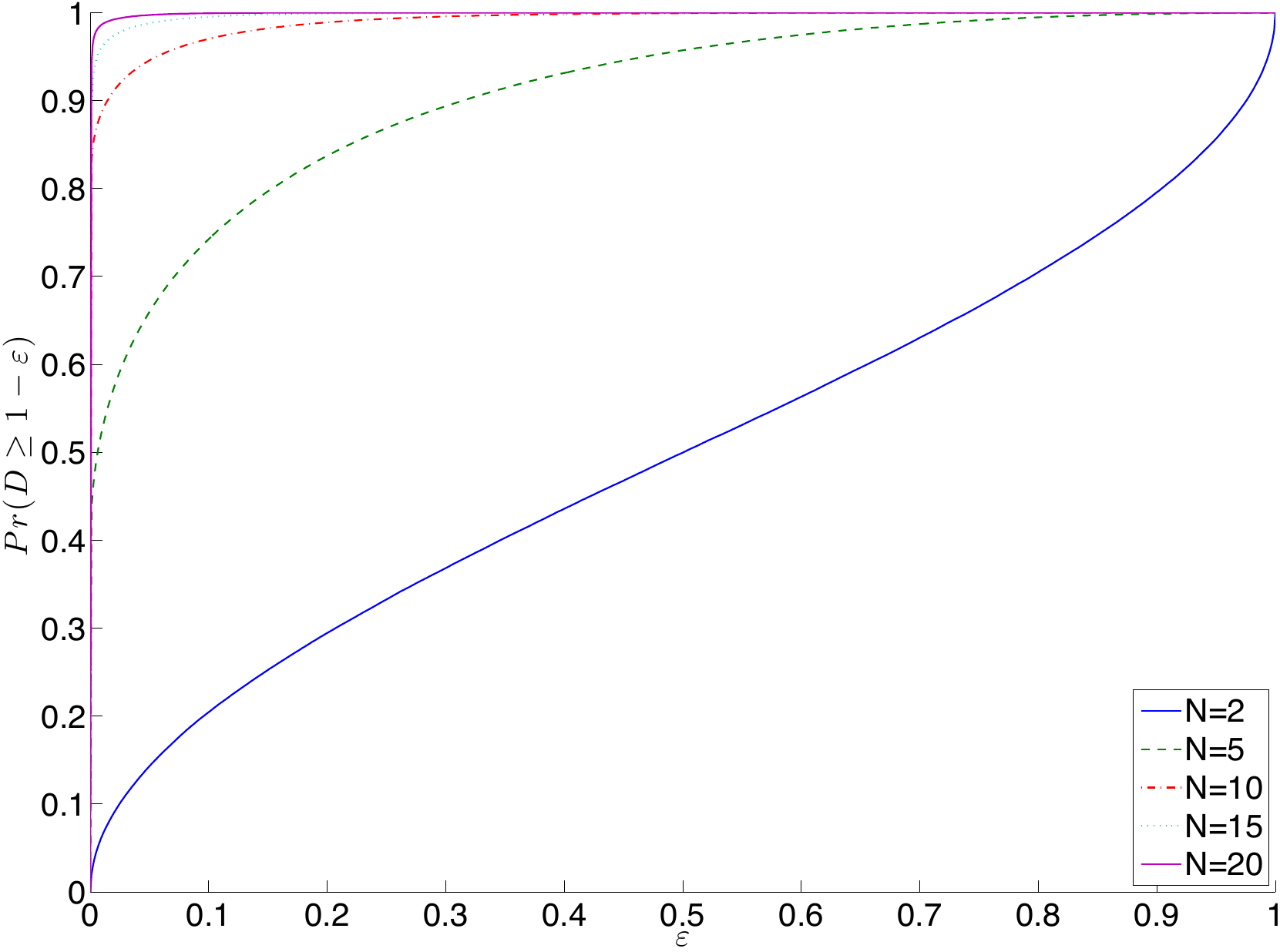}
\caption{(color online) The probability of picking a state with maximum distinguishability $D \geq 1-\epsilon$ as a function of $\epsilon$ for different dimensions $N$.}
\label{fig:pdatom}
\end{figure}

We can also again determine the probability that a randomly chosen state has a maximum distinguishability greater than $1-\epsilon$. This is shown in Fig.~\ref{fig:pdatom}. This figure confirms that the maximum distinguishability tends to be higher for atomic systems than for harmonic systems.

Finally, in Fig.~\ref{fig:avDatom} we show the average maximum distinguishability over the quantum state space as a function of the dimensionality of the system. Included in the figure are the standard deviations above and below the average. The distinguishability $\braket{D}$ approaches unity faster for atomic systems than for harmonic systems, but the fluctuations in $D$ are also larger for atomic systems compared to harmonic systems. 

\section{Implications for quantum speed limits}\label{sec:qsl}\noindent
The original Mandelstam-Tamm bound is easily expressed in terms of the distinguishability $D$, or rather the deviation from orthogonality $\eta$ (we write $\eta$ instead of $\epsilon$ because we consider a particular value for $D$, and $\epsilon$ denotes our distinguishability \emph{threshold}):
\begin{align}
 \tau \geq \frac{\hbar}{\Delta E} \arccos\sqrt{\eta} \simeq \frac{\pi}{2} \frac{\hbar}{\Delta E} \left( 1-\frac{2\sqrt{\eta}}{\pi} \right)\, ,
\end{align}
where the approximation is valid for small $\eta$.

The fact that systems do not in general evolve to orthogonal states has repercussions for physical properties that rely on this assumption. As an important example we consider the Margolus-Levitin bound on the speed of dynamical evolution \cite{margolus98}. Margolus and Levitin define the quantity $S = \braket{\psi|\psi(t)}$ and derive the inequality
\begin{align}
 \Re(S) \geq 1-\frac{2E}{\pi\hbar} + \frac{2}{\pi} \Im(S)\, ,
\end{align}
where $E$ is the average energy of the system above the ground state. By requiring that $\ket{\psi(t)}$ is orthogonal to $\ket{\psi}$ the real and imaginary parts of $S$ must be zero, and as a result we obtain the bound
\begin{align}\label{eq:ml}
 t \geq \frac{\pi}{2} \frac{\hbar}{E}\, .
\end{align}
This is commonly interpreted as the minimum time it takes for a system to evolve to a distinct quantum state. However, if $S$ is never zero this bound must be modified. From $\abs{S}^2 = 1-D=\eta$ we deduce that $\Re(S)^2 \geq \eta$ and $\Im(S)^2 \geq \eta$. In turn, this produces the bound
\begin{align}
 \Abs{\Re(S) - \frac{2}{\pi} \Im(S)} \leq \sqrt{\eta}\, .
\end{align}
This leads to a modified Margolus-Levitin bound 
\begin{align}\label{eq:mmlbound}
 t \geq \frac{\pi}{2} \frac{\hbar}{E} (1-\sqrt{\eta})\, ,
\end{align}
which now depends on the level of orthogonality $\eta$, and by extension on the type of system (harmonic, atomic, etc.). In other words, the bound no longer depends only on the average energy above the ground state. This is also tighter bound than Eq.~(\ref{eq:ml}), although for optimal states $\eta\to0$ the bounds coincide.

\begin{figure}[t!]
\includegraphics[width=8.5cm]{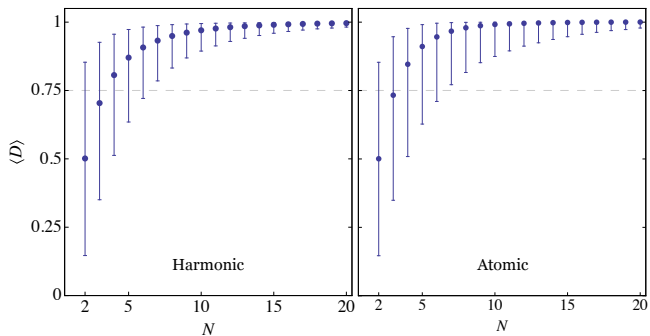}
\caption{(color online) The average maximum distinguishability $\braket{D}$ as a function of the dimensionality of the harmonic (left plot) and atomic (right plot) systems, respectively. Note how $\braket{D}$ approaches 1 faster for atomic systems, but the fluctuations in $D$ are also larger for atomic systems compared to harmonic systems. The dashed horizontal line is there to assist the eye in the comparison of the two graphs.}
\label{fig:avDatom}
\end{figure}

We can view $1-\eta$ as the average maximum distinguishability, corresponding to the points in Fig.~\ref{fig:avDatom}. However, the modified Margolus-Levitin bound in Eq.~(\ref{eq:mmlbound}) is then an \emph{average} bound, and there will always be a (small) probability that the bound is violated when we pick a random state. This is a particularly important effect in lower dimensions. The modified bound was derived in Ref.~\cite{soderholm99} from a direct construction of the optimal states given $\eta$.

\section{Conclusions}\label{sec:con}
\noindent
In this paper, we have studied the dynamical speed of evolution and the average attainable maximum distinguishbility of quantum systems in Hilbert spaces of dimension up to $N=20$. We found that the details of the dynamics (in the form of the Hamiltonian) not only determine the speed of dynamical evolution, which is well known, but it also determines the level of distinguishability. Systems with irregular energy spectra evolve on average to more distinguishable states than systems with a regular energy spectrum, but they also are likely to take longer to do so. The Mandelstam-Tamm and Margolus-Levitin bounds are easily modified to take this low-dimensional behaviour into account.

\section*{Acknowledgments}\noindent
We thank Norman Margolus and David Whittaker for stimulating discussions.

\appendix

\section{Distinguishability threshold in $N=2$}\label{app:pop}\noindent
The probability that for a random state in a two-dimensional state space the maximally distinguishable state has a value of $D\geq 1-\epsilon$ can be decomposed into 
\begin{align}
 \pr{D\geq 1-\epsilon} = \sum_{\mathbf{x}} \pr{D\geq 1-\epsilon|\mathbf{x}} \pr{\mathbf{x}}\, ,
\end{align}
where $\mathbf{x} = (x_1,x_2)$. The conditional probability inside the summation over $\mathbf{x}$ is a Heaviside function:
\begin{align}
 \pr{D\geq 1-\epsilon|\mathbf{x}} = \Theta(D(\mathbf{x}) - 1+\epsilon)\, .
\end{align} 
For $\pr{\mathbf{x}}$ we choose a uniform distribution $\exp(-r^2)/\pi$, and using Eq.~(\ref{eq:dist}) we find 
\begin{align}
  \pr{D\geq 1-\epsilon} = \frac{1}{\pi} \int_{\mathbb{R}^2} d\mathbf{x}\; \Theta\left( \epsilon-\Abs{\frac{x_1^2-x_2^2}{x_1^2+x_2^2}}^2 \right) e^{-x_1^2-x_2^2} \, .
\end{align}
We evaluate this integral by manipulating the domain of integration. First, by inspecting the symmetries of the integrand we note that we can rewrite the integral as 
\begin{align}
  \pr{D\geq 1-\epsilon} = \frac{4}{\pi} \int_0^\infty\int_0^\infty d\mathbf{x}\; \Theta\left( \epsilon-\Abs{\frac{x_1^2-x_2^2}{x_1^2+x_2^2}}^2 \right) e^{-x_1^2-x_2^2} \, .
\end{align}
We write the argument of the Heaviside function as
\begin{align}
 x_1^4 (1-\epsilon) - 2 x_1^2 x_2^2 (1+\epsilon) + x_2^4 (1-\epsilon) \leq 0\, ,
\end{align}
and solve for $x_1^2$:
\begin{align}
 {x_1^2}_\pm = x_2^2 \left( \frac{2}{1\pm\sqrt{\epsilon}}-1 \right) .
\end{align}
This leads to the new limits of integration for $x_1$:
\begin{align}
 {x_1}_\pm = x_2 \left( \frac{2}{1\pm\sqrt{\epsilon}}-1 \right)^\frac12 \equiv x_2 \alpha_\pm \, ,
\end{align}
yielding the double integral
\begin{align}
 \pr{D\geq 1-\epsilon} = \frac{4}{\pi} \int_0^\infty dx_2 \int_{x_2 \alpha_-}^{x_2 \alpha_+} dx_1\; e^{-x_1^2-x_2^2} \, .
\end{align}
Converting to polar coordinates gives
\begin{align}
 \pr{D\geq 1-\epsilon} = \frac{2}{\pi} \int_{\phi_-}^{\phi^+} \int_0^\infty dr\, d\phi\;  r e^{-r^2} \, ,
\end{align}
with $\phi_\pm = \arctan\alpha_\pm$. From this, the result in Eq.~(\ref{eq:n2eps}) follows immediately.

\end{document}